\documentclass[aps,prb,twocolumn,showpacs]{revtex4}
\usepackage{graphicx}
\usepackage{dcolumn}
\usepackage{bm}
\usepackage{stmaryrd}
\usepackage{latexsym}
\usepackage{amssymb}
\usepackage{amsfonts}
\usepackage{amsmath}

\begin{document}

\title{Effect of edge decoration on the energy spectrum of semi-infinite lattices}

\author{Yuanyuan Zhao, Wei Li and Ruibao Tao}
\affiliation{State Key Laboratory of Surface Physics and Department
of Physics,\\ Fudan University, Shanghai 200433, People's Republic of China}

\date{\today}

\begin{abstract}
Analytical studies of the effect of edge decoration on the energy
spectrum of semi-infinite one-dimensional (1D) lattice chain with
Peierls phase transition and zigzag edged graphene (ZEG) are
presented by means of transfer matrix method, in the frame of which
the sufficient and necessary conditions for the existence of the
edge states are determined. For 1D lattice chain, the zero-energy
edge state exists when Peierls phase transition happens regardless
whether the decoration exists or not, while the non-zero-energy edge
states can be induced and manipulated through adjusting the edge
decoration. On the other hand, the semi-infinite ZEG model with
nearest-neighbor interaction can be mapped into the 1D lattice chain
case. The non-zero-energy edge states can be induced by the
decoration as well, and we can obtain the condition of the
decoration on the edge for the existence of the novel edge states.
\end{abstract}

\pacs{73.22.Pr,73.20.At,71.15.-m} \maketitle

\section{Introduction}
\label{intro}

One of the interesting phenomena in condensed matter physics is the
existence of edge states, the properties of which are distinct from
those of the bulk states. There are some examples showing that the
system is insulated in the bulk, while conduction can be allowed by
edge states on the boundary. The most prominent ones are the quantum
Hall effect (QHE)\cite{QHE, LaughlinQHE, HalperinQHE, Bhatt} and the
quantum spin Hall effect (QSHE)\cite{Kane1, Bernevig1, Bernevig2},
where the quantization of a Hall conductance is tightly associated
with the edge states\cite{LaughlinQHE, HalperinQHE, Bhatt, Kane1,
Bernevig1, Bernevig2, Thouless, Kohmoto, Hatsugai1, Qi1, Kane2,
Kane3}.

With the development of graphene\cite{Novoselov1, Novoselov2} in
recent years, the edge states of graphene model\cite{Castro} attract
much attention. From the topological view, edge states can be
induced in the system with different structures of edges\cite{Klein,
Fujita}; in experiments, with the help of scanning tunneling
microscopy and scanning tunneling spectroscopy, the presence of
structure-dependent edge states of graphite can be
observed\cite{Niimi,Kobayashi}. As we know, some results of the edge
states have been reported in the tight-binding model\cite{Fujita,
Nakada, Wakabayashi, Kusakabe, Peres, Ezawa, Hasegawa1, Hasegawa2},
Dirac equation\cite{Brey, Sasaki, Abanin}, the transfer matrix
method (TMM) calculation\cite{DHLee, LJiang, Haidong, Li-Tao}, first
principles calculations\cite{Miyamoto, Okada, Lee}, and most of them
focus on the zero mode of the graphene nanoribbon, while some
consider about the effects of disorder or the edge corrections.

In this paper, we report a systematic study of the zero-energy and
non-zero-energy edge states caused by the decoration on the edge by
means of TMM. We analytically study the simple tight-binding model
of semi-infinite 1D lattice chain and the zigzag edged graphene
(ZEG), and the consequent novel edge states that arise due to the
edge decoration, which is the hopping interaction on the edge being
different from that in the bulk, caused as edge binding softening or
stiffening. The extended states and the localized edge states are
also discussed, and the analytical relations of the edge states with
the edge decoration can be obtained as well. It might be applicable
in future devices since novel edge states, particularly in the
energy forbidden region, could significantly change the optical
property, even the transportation.

The rest of this paper is organized as follows. In Sec. \ref{1Dsec},
we take a semi-infinite 1D Peierls chain model as a fundamental case
to show the relation of edge states with phase transition and
describe how the analytical relations can be obtained efficiently
through TMM for not only the extended states, but also the edge
states. The sufficient and necessary conditions (SNC) for the
existence of the edge states are determined. With the general
discussion of the 1D chain model, in Sec. \ref{Zigsec}, we map it to
the ZEG, with applying the rigorous results in Sec. \ref{1Dsec} to
obtain the edge states of ZEG with and without edge decoration.
Finally, we give a brief conclusion in Sec. \ref{conclusions}.

\section{EDGE STATES AND PEIERLS PHASE TRANSITION}\label{1Dsec}

\begin{figure}[h]
\centering
\includegraphics[scale=0.7]{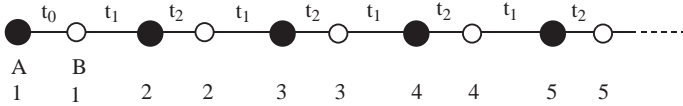}
\caption{Schematic illustration of the lattice structure of 1D
semi-infinite chain with the closed (black) circles and the open
(white) circles, consisting of sublattices $A$ and
$B$.}\label{1dchainpic}
\end{figure}

As a typical example for studying edge states occurring in
semi-infinite systems due to phase transition, we treat a
semi-infinite 1D lattice chain that can occur Peierls phase
transition. It has been well-known that 1D periodic atomic chain is
not stable due to the electron-phonon interaction that drives chain
to have pairing transition. In the case of half-filling electrons,
the atoms in a periodic lattice chain like to pair together, forming
Peierls pairing state. Its geometrical structure is shown in Fig.
\ref{1dchainpic}, where one unit cell includes two atoms. In the
tight-binding approximation, consider just the nearest-neighbor
interaction (n.n.), two different n.n. hopping constants are
$t_{1}(=t\pm \Delta )$ and $t_{2}(=t\mp \Delta )$ separately, where
$t $ is the hopping constant before Peierls phase transition,
$\Delta (\geqslant 0)$ describes the pairing phase transition just
like the phase transition in 1D molecule polyacetylene. $\Delta =0$
when Peierls phase transition is suspended. In general, $t_{1}\neq
t_{2}$ if $\Delta \neq 0.$ The chain with Peierls pairing is
equivalent to a 1D model of atomic chain in which each cell has two
atoms, which are denoted in a cell as $A$ and $B$. The edge of the
semi-infinite chain is set at the left terminal. Without loss of
generality, we assume that the lattice at left terminal is $A$
denoted as $1$, and label the unit cell with an integer number
ordered from the left terminal to the right infinity. Since a
non-Bravais lattice with two atoms per unit cell is bipartite, we
have defined two kinds of fermion operators, $\varphi_{iA}$ and
$\varphi_{iB}$, where $i$ is the integer index labeling the unit
cell. The Hamiltonian can be approximated by

\begin{equation}
H=t_{0}\varphi_{1A}^{\dag}\varphi_{1B}+t_{1}\sum_{i\geqslant
1}\varphi_{i+1A}^{\dag}\varphi_{iB}+t_{2}\sum_{i\geqslant
2}\varphi_{iA}^{\dag}\varphi_{iB}+h.c.  \label{1dham}
\end{equation}
where $t_{0},t_{1},t_{2}$ are three hopping parameters, and set all
of them to be positive. The parameter $t_{0}$ is introduced to
describe edge decoration. From Eq. (\ref{1dham}), we can derive the
dynamical equations\cite{Harper} through defining an elemental
excitation with energy $E$ as follows

\begin{equation}
\begin{cases}
E\varphi _{1A} & =\quad t_{0}\varphi _{1B} \\
E\varphi _{1B} & =\quad t_{0}\varphi _{1A}+t_{1}\varphi _{2A} \\
E\varphi _{iA} & =\quad t_{2}\varphi _{iB}+t_{1}\varphi _{i-1B}\quad
(i\geqslant 2) \\
E\varphi _{iB} & =\quad t_{2}\varphi _{iA}+t_{1}\varphi _{i+1A}\quad
(i\geqslant 2)
\end{cases}
\label{1drec}
\end{equation}
According to above equations, we try to get the bulk wave-functions,
and find out the physically meaningful localized states which decay
from the surface. When $E=0$ or $E\neq 0$, the above equations will
be in different forms, so in the following subsection, we will
discuss them separately.\\
\newline
Under the condition $E=0$, Eq. (\ref{1drec}) can be reduced to the
decoupled ones:
\begin{equation}\label{1D-zero}
\begin{cases}
\qquad \qquad t_{0}\varphi _{1B} & =\quad 0 \\
t_{0}\varphi _{1A}+t_{1}\varphi _{2A} & =\quad 0 \\
t_{2}\varphi _{iB}+t_{1}\varphi _{i-1B} & =\quad 0\quad (i\geqslant 2) \\
t_{2}\varphi _{iA}+t_{1}\varphi _{i+1A} & =\quad 0\quad (i\geqslant
2)
\end{cases}
\end{equation}
From Eq. (\ref{1D-zero}), it is easy to know that the wave-functions
of all the type $B$ sublattices will be equal to zero: $\{\varphi
_{iB}=0:i=1,2,\cdots ,\infty \}$; meanwhile the wave-function of
sublattice $A$ in the $n$ unit cell will be:
\begin{equation}
\varphi
_{nA}=\left(-\frac{t_{2}}{t_{1}}\right)^{n-2}\left(-\frac{t_{0}}{t_{1}}\right)\varphi
_{1A},\qquad n=2,3,4\cdots
\end{equation}
$\varphi _{1A}$ is the initial value of wave excitation with zero
energy at edge site. $t_{2}=t_{1}$ corresponds to the extended state
in bulk. If $t_{1} \neq t_{2}$, in order to obtain the physically
meaningful edge states, we must restrict $t_{2}/t_{1}<1 $, so that
when $n$ goes to infinity, the wave-functions of sublattice $A$ will
approach to $0$, showing the state is localized near the edge
region. Away from the surface, the amplitude of the wave-function
will exponentially decay as $ e^{-n2a/\lambda _{L}}$. The localized
length $\lambda _{L}$ of the edge state is $2a/(\ln t_{2}-\ln
t_{1})$, where $2a$ is the periodic constant of the cell after
Peierls pairing. Thus, we can conclude that zero-energy edge state
can exist due to Peierls phase transition when $t_{2}<t_{1}$
$(\Delta \neq 0)$, and this result can be obtained by surface
Green's function\cite{Peres1}. It is shown that no matter whether
the edge decoration exists or not, zero-energy edge state is not
affected by edge decoration $t_{0}(\neq t_{2})$. 1D Peierls pairing
mode is an example to demonstrate that the studies of zero-energy
edge state could show some evidence of the phase transition, even
the mechanism of phase transition.

In the following, we will study whether the non-zero-energy edge
states can exist or not. According to Eq. (\ref{1drec}), after
eliminate sublattice $A$ and $B$ separately, we can obtain

\begin{equation}
\begin{cases}
E^{2}\varphi_{1A}= t_{0}^{2}\varphi_{1A}+t_{0}t_{1}\varphi _{2A} \\
E^{2}\varphi_{2A}=(t_{1}^{2}+t_{2}^{2})\varphi_{2A}+t_{0}t_{1}\varphi
_{1A}+t_{1}t_{2}\varphi _{3A}\\
E^{2}\varphi_{iA}=(t_{1}^{2}+t_{2}^{2})\varphi_{iA}+t_{1}t_{2}\varphi_{i-1A}+t_{1}t_{2}\varphi
_{i+1A}, i\geqslant 3,
\end{cases} \label{1dAtr}
\end{equation}
and
\begin{equation}
\begin{cases}
E^{2}\varphi_{1B}=(t_{0}^{2}+t_{1}^{2})\varphi_{1B}+t_{1}t_{2}\varphi_{2B}, \\
E^{2}\varphi_{iB}=(t_{1}^{2}+t_{2}^{2})\varphi_{iB}+t_{1}t_{2}\varphi_{i-1B}
+t_{1}t_{2}\varphi_{i+1B},i\geqslant
2
\end{cases}
\label{1dBtr}
\end{equation}
According to the first equation in Eq. (\ref{1dAtr}), we introduce a
fictitious $\varphi _{0A}=0$ into our discussion without lose of
generality. Then we can turn Eq. (\ref{1dAtr}) into matrix forms:

\begin{eqnarray}
\left(
\begin{array}{c}
\varphi _{i+1A} \\
\varphi _{iA}
\end{array}
\right)  &=& T^{i-2}\left(
\begin{array}{c}
\varphi _{3A} \\
\varphi _{2A}
\end{array}
\right) \quad (i\geqslant 3)  \notag \\
\left(
\begin{array}{c}
\varphi _{3A} \\
\varphi _{2A}
\end{array}
\right)  &=&T_{1}\left(
\begin{array}{c}
\varphi _{2A} \\
\varphi _{1A}
\end{array}
\right)   \notag \\
\left(
\begin{array}{c}
\varphi _{2A} \\
\varphi _{1A}
\end{array}
\right)  &=&T_{2}\left(
\begin{array}{c}
\varphi _{1A} \\
\varphi _{0A}
\end{array}
\right)
\end{eqnarray}
where $T$, $T_{1}$ and $T_{2}$ are defined as
\begin{eqnarray}
T &=&\left(
\begin{array}{cc}
\alpha  & -1 \\
1 & 0
\end{array}
\right)    \\
T_{1} &=&\left(
\begin{array}{cc}
\alpha  & -\frac{t_{0}}{t_{2}} \\
1 & 0
\end{array}
\right)   \notag \\
T_{2} &=&\left(
\begin{array}{cc}
\beta  & -1 \\
1 & 0
\end{array}
\right)   \notag
\end{eqnarray}
with $\alpha =(E^{2}-t_{1}^{2}-t_{2}^{2})/t_{1}t_{2}$ and $\beta
=(E^{2}-t_{0}^{2})/t_{0}t_{1}.$ Meanwhile, the wave-functions of the
sublattices $B$ satisfy the relations:

\begin{eqnarray}
\left(
\begin{array}{c}
\varphi _{i+1B} \\
\varphi _{iB}
\end{array}
\right)  &=& T^{i-1}\left(
\begin{array}{c}
\varphi _{2B} \\
\varphi _{1B}
\end{array}
\right) \quad ,(i\geqslant 2);  \notag \\
\varphi _{2B} &=&\gamma \varphi _{1B};\varphi
_{1B}=\frac{E}{t_{0}}\varphi _{1A}
\end{eqnarray}
and $\gamma =(E^{2}-t_{0}^{2}-t_{1}^{2})/t_{1}t_{2}$. The energy $E$
must be real for the physical quantity, and all $\alpha ,\beta $ and
$ \gamma $ are real numbers. Diagonalize the transfer matrix $T$ to
$ D=U^{-1}TU$, and we have

\begin{eqnarray}\label{states}
\begin{cases}
\left(
\begin{array}{c}
\varphi _{i+1A} \\
\varphi _{iA}
\end{array}
\right)=UD^{i-2}U^{-1}T_{1}T_{2}\left(
\begin{array}{c}
\varphi _{1A} \\
\varphi _{0A}
\end{array}
\right) \\
\left(
\begin{array}{c}
\varphi _{i+1B} \\
\varphi _{iB}
\end{array}
\right)=UD^{i-1}U^{-1}\left(
\begin{array}{c}
\gamma  \\
1%
\end{array}
\right) \frac{E}{t_{0}}\varphi _{1A}.
\end{cases}
\end{eqnarray}
with
\begin{equation}
D=\left(
\begin{array}{cc}
\lambda _{-} & 0 \\
0 & \lambda _{+}
\end{array}
\right) \qquad U=\left(
\begin{array}{cc}
\lambda _{-} & \lambda _{+} \\
1 & 1%
\end{array}
\right)   \label{1-Dtransform}
\end{equation}
\begin{equation}\label{inverseU}
U^{-1}=\frac{1}{\sqrt{\alpha ^{2}-4}}\left(
\begin{array}{cc}
-1 & \lambda _{+} \\
1 & -\lambda _{-}
\end{array}
\right)
\end{equation}
where $\lambda _{-}=\frac{1}{2}(\alpha -\sqrt{\alpha ^{2}-4})$ and
$\lambda _{+}=\frac{1}{2}(\alpha +\sqrt{\alpha ^{2}-4})$.
Eq. (\ref{states}) can be rewritten as
\begin{equation}\label{1D-trans-2}
\begin{cases}
\left(
\begin{array}{c}
\varphi _{n+3A} \\
\varphi _{n+2A}
\end{array}
\right) =U\left(
\begin{array}{c}
\Gamma _{1}\lambda _{-}^{n} \\
\Gamma _{2}\lambda _{+}^{n}
\end{array}
\right) \varphi _{1A},  \\
\left(
\begin{array}{c}
\varphi _{n+3B} \\
\varphi _{n+2B}
\end{array}
\right) =U\left(
\begin{array}{c}
\Theta _{1}\lambda _{-}^{n} \\
\Theta _{2}\lambda _{+}^{n}
\end{array}
\right) \dfrac{E}{t_{0}}\varphi _{1A}.
\end{cases}
\end{equation}
Where $\Gamma _{1}=U_{11}^{-1}\left( \alpha \beta
-\frac{t_{0}}{t_{2}} \right) +U_{12}^{-1}\beta $, $\Gamma
_{2}=U_{21}^{-1}\left( \alpha \beta - \frac{t_{0}}{t_{2}}\right)
+U_{22}^{-1}\beta $, $\Theta _{1}=U_{11}^{-1}\gamma +U_{12}^{-1}$,
$\Theta _{2}=U_{21}^{-1}\gamma +U_{22}^{-1}$. We must request
$\varphi _{1A}\neq 0,$ otherwise the wave function vanishes
$\{\varphi _{iA}=\varphi _{iB}=0:i=1,2,\cdot \cdot \cdot ,\infty
\}$.

When $|\alpha |\leqslant 2$, $|\lambda _{-}|=|\lambda _{+}|=1 $. It
corresponds to the extended states in the bulk. Thus, all energy
$E^2$ satisfying the energy inequality
$|E^2-t^2_1-t^2_2|/|t_1t_2|\leqslant 2$ must associate with the
extended states. With the Eq. (\ref{1dAtr}) and Eq. (\ref{1dBtr}),
considering a continuous parameter $k_x$ to keep $|\cos(k_xa)|
\leqslant 1$, we can get the dispersion relation of the bulk states
as:

\begin{equation}\label{1D-dispersion}
E^2=t_1^2+t_2^2+2t_1t_2 \cos(k_xa), ~k_xa\in [-\pi,\pi].
\end{equation}
In the case, we have $\lambda_{\pm}=e^{\pm ik_xa}$. From
Eq. (\ref{1D-trans-2}), the $k_x$ belonging to $\mathbb{R}$ is
clearly served as a wave vector of this extended mode propagating
along the chain direction. The left edge makes the semi-infinite
chain to have two waves interfered. One is the $e^{ik_xna}$, the
other is $e^{-ik_xna}$.

In the following, we focus on the solution of edge state when the
condition $|\alpha |>2$ is satisfied. In the case of $\alpha
>2$, we get $0<\lambda _{-}<1$ and $\lambda _{+}>1$, while $\lambda
_{-}<-1$ and $-1<\lambda _{+}<0$ for the case of $\alpha <-2$. We
will discuss these two cases respectively.

In the case of $\alpha >2$, it yields  $0<\lambda _{-}<1$, $\lambda
_{+}>1$. The necessary conditions to have a stable edge states
localized at edge boundary must be $\Gamma _{2}=0$ and $\Theta
_{2}=0$ that are

\begin{eqnarray*}
U_{21}^{-1}\left( \alpha \beta -\frac{t_{0}}{t_{2}}\right)
+U_{22}^{-1}\beta
&=&0 \\
U_{21}^{-1}\gamma +U_{22}^{-1} &=&0
\end{eqnarray*}
From Eq. (\ref{1-Dtransform}), Eq. (\ref{inverseU}) and the
definitions of $\lambda_{\pm}$, above two equations can be changed
to the following:
\begin{eqnarray}
\begin{cases}
(\alpha +\sqrt{\alpha ^{2}-4})\beta -2\dfrac{t_{0}}{t_{2}} =0, \\
\sqrt{\alpha ^{2}-4}-\alpha +2\gamma  =0  \label{gamma}
\end{cases}
\end{eqnarray}
where $\alpha, \beta$ and $\gamma$ are the functions of energy
$E(\neq 0)$. Eq. (\ref{gamma}) is the necessary condition for the
occurrence of non-zero-energy edge state. If Eq. (\ref{gamma}) can
not be satisfied for any finite energy $E$, the non-zero-energy edge
states could not exist. Inputting the definitions of $\alpha, \beta$
and $\gamma$ into Eq. (\ref{gamma}), we obtain

\begin{equation*}
(t_{0}^2-t_{2}^2)(E^2-t_{0}^2)=t_{0}^2t_{1}^2
\end{equation*}
When the edge has no decoration, $t_{0}=t_{2}(\neq 0)$, there should
be no finite solution of $E$ for above equation since
$t_{0}^2t_{1}^2\neq 0$. Thus we conclude that non-zero-energy edge
state does not exist in 1D Peierls chain without edge decoration.
When $t_{0}\neq t_{2}$, we have solution
\begin{equation}
E^{2}=t_{0}^{2}+t_{0}^{2}t_{1}^{2}/\left(
t_{0}^{2}-t_{2}^{2}\right).
\end{equation}
Due to $\alpha >2$, we obtain $ E^{2}>(t_{1}+t_{2})^{2}>0$, which
corresponds to a stable edge state. With the definition of $\gamma $
and $\alpha $, we get $\gamma =\alpha -\left(
t_{0}^{2}-t_{2}^{2}\right) /t_{1}t_{2}$. Taking it back to Eq.
(\ref{gamma}), we have
\begin{equation*}
\left( t_{0}^{2}-t_{2}^{2}\right) /t_{1}t_{2}=\frac{\alpha
}{2}+\frac{\sqrt{ \alpha ^{2}-4}}{2}>1
\end{equation*}
Thus the condition to get an additional stable non-zero-energy edge
state is the inequality $t_{0}^{2}>t_{2}^{2}+t_{1}t_{2}$. An
appropriate manipulation on the edge decoration $t_{0}$ can induce
the non-zero-energy edge state. There is no constraint showing the
quantity relation between $t_{1}$ and $t_{2}$. From previous
subsection, it has been known that the zero-energy edge state can
exist with $t_{2}/t_{1}<1$, so the zero mode and the non-zero-energy
edge states can appear at the same time. However, there will be
non-zero-energy edge states only if $t_{2}\geqslant t_{1}$ together
with $ t_{0}^{2}>t_{2}^{2}+t_{1}t_{2}$.

Similarly we can get constraints for $\alpha <-2$
\begin{eqnarray*}
(\alpha -\sqrt{\alpha ^{2}-4})\beta -2\frac{t_{0}}{t_{2}} &=&0 \\
\sqrt{\alpha ^{2}-4}+\alpha -2\gamma  &=&0
\end{eqnarray*}
They also yield the same energy condition: $
E^{2}=t_{0}^{2}+t_{0}^{2}t_{1}^{2}/\left( t_{0}^{2}-t_{2}^{2}\right)
$. And besides $\gamma =\alpha -\left( t_{0}^{2}-t_{2}^{2}\right)
/t_{1}t_{2}$, when $\alpha <-2$, it is easy to get:
\begin{equation*}
\left( t_{0}^{2}-t_{2}^{2}\right) /t_{1}t_{2}=\frac{\alpha
}{2}-\frac{\sqrt{ \alpha ^{2}-4}}{2}<-1
\end{equation*}
We can find that the non-zero-energy edge state can also exist under
the condition: $t_{0}^{2}<t_{2}(t_{2}-t_{1})$ that requests
$t_{2}>t_{1}$ and results zero-energy edge state vanishing.

From above two cases, we can see that non-zero-energy edge states
are tightly related with edge decoration $t_{0}$ and can be
manipulated by decorating the edge atoms.

We conclude that a zero-energy edge state will appear if Peierls
transition happens $ (t_{2}<t_{1})$ regardless edge decoration at
all. A non-zero-energy edge state can occur due to some edge
decoration as following condition:
\begin{equation}\label{1D-edge}
\begin{cases}
E^{2}=t_{0}^{2}+t_{0}^{2}t_{1}^{2}/\left(
t_{0}^{2}-t_{2}^{2}\right) (>0,t_{0}\neq t_{2}): \\
t_{0}^{2}>t_{2}^{2}+t_{1}t_{2}>0;~ \text{or}
~t_{0}^{2}<t_{2}(t_{2}-t_{1}),~ t_{2}>t_{1}.
\end{cases}
\end{equation}
Non-zero-energy edge state can be manipulated with decorating
the edge.

\section{EDGE STATES OF SEMI-INFINITE ZEG}\label{Zigsec}

In this section, we would present some analytical results for the
edge states of ZEG. Although some results has been
reported\cite{LJiang} for uniform ZEG ribbon, and can be extended to
semi-infinite ZEG through size scaling to the infinite width of
ribbon, the effect of edge decoration on the edge states has not
been reported. Thus we will supply some analytical results on the
occurrence of non-zero-energy edge states and its energy dispersion
in semi-infinite ZEG with edge decoration. In the previous section,
we have obtained some criteria on the existence of edge states in
semi-infinite 1D Peierls model with and without edge decoration. In
fact, some 2D and 3D lattice models could be reduced into above 1D
model such as the model of ZEG to be discussed in this section. The
geometrical structure of graphene is shown in Fig.
\ref{semi-graphene}. Without loss of generality, we denote the the
first left line $1_{A}$ as an edge, and take the lattice system in
$y$ direction to be infinite and periodic with constant $a$. Then
the semi-infinite ZEG can be described by a line set with infinite
number of periodic 1D perpendicular chain along $y$ direction:
$\{1_{A},1_{B},...,n_{A},n_{B},\cdots\}.$ The positions of
sublattices $A$ and $B$ are labeled by two indices $
\{i,y_{i,A(B)}\}$ where $i$ labels the number of perpendicular line
from $1$ to $\infty$. After Fourier transformation for
$\{y_{i,A(B)}\}$ with periodic constant $a,$ in second quantization
representation, the model Hamiltonian with n.n. hopping can be
expressed by a set of the Fermion operators $\{\phi
_{i,A(B)}(k_{y}):i=1,2,...,\infty \}$:
\begin{figure}[h]
\centering\includegraphics[width=0.45\textwidth]{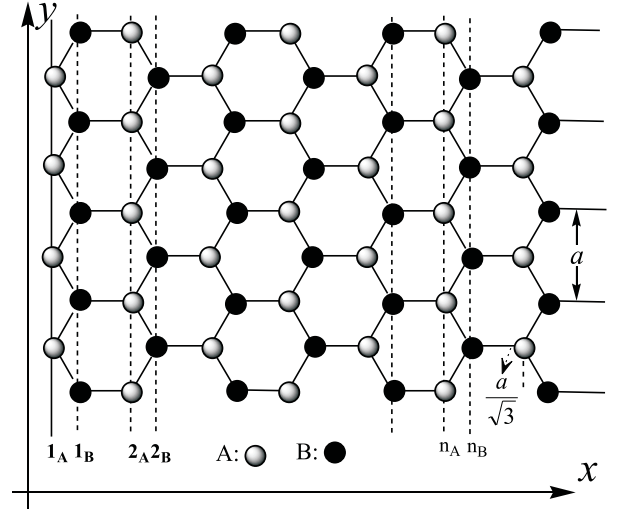}
\caption{Schematic
illustration of the lattice structure of semi-infinite zigzag edged graphene,
the fulfilled circles represent the sublattice $B$, others are the sublattice
$A$. The left first line $1_A$ is the zigzag edge of the graphene. }
\label{semi-graphene}
\end{figure}

\begin{eqnarray}
H &=&\sum_{k_{y},i\geqslant 2}t \phi_{i,A}^{\dag}\left( k_{y}\right)
\phi _{i-1,B}(k_{y}) +\tilde{t}\phi_{i,A}^{\dag}\left( k_{y}\right)
\phi_{i,B}\left( k_{y}\right)   \notag \\
&+&\sum_{k_{y}}\tilde{t}_0\phi_{1,A}^{\dag}(k_{y})\phi
_{1,B}(k_{y})+h.c., \label{zigzagH}
\end{eqnarray}
where $k_{y}$ is a good quantum number, $\tilde{t}=2t\cos
(k_{y}a/2)$ and $\tilde{t}_0=2t'_{0}\cos (k_{y}a/2)$. We have
introduced a simplified edge hopping decoration $t'_{0}$ at first
line. When $t'_{0}=t,$ it becomes an ideal semi-infinite ZEG without
edge decoration. In general, $t'_{0}\neq t$ that describes some edge
decoration, and we set all the hopping interaction positive as well.
The wave function at site $(i,j)$ of $A(B)$ sublattice is written as
$\Psi_{A(B)}^{(i,j)}(k_{y})=\exp
(ik_{y}y_{j,A(B)})\phi_{i,A(B)}(k_y)$.

For each fixed $k_y$, the Hamiltonian of Eq. (\ref{zigzagH}) is
equivalent to the 1D Peierls chain in Sec. \ref{1Dsec} when we
denote $t_{0}=\tilde{t}_0$, $t_{1}=t$ and $t_{2}=\tilde{t}$. Thus
the conclusion in Sec. \ref{1Dsec} can be applied to study the ZEG.

Comparing with standard energy spectrum of bulk graphene, $\alpha =
\frac{E^{2}-4 t^{2}\cos^{2}{k_{y}a/2}-t^{2}}{2t^{2}\cos{k_{y}a/2}}$,
when $\alpha^2\leqslant 4$, we can get the dispersion relation for
extended states:
\begin{eqnarray}\label{ZEG-extend}
E^{2}&=&t^{2}_{1}+t^{2}_{2}+2t_{1}t_{2}\cos(\sqrt{3}k_{x}a/2)\notag\\
&=&t^2[1+4\cos^2(k_{y}a/2)+4\cos(k_{y}a/2)\cos(\sqrt{3}k_{x}a/2)]\notag\\
\end{eqnarray}
This is the standard formula of the energy spectrum of the infinite
bulk graphene.

In the following, we would focus on the edge states. From Sec.
\ref{1Dsec}, the necessary condition for the existence of
zero-energy edge state is $|t_{2}/t_{1}|<1$ that directly results
the condition $|2\cos(k_{y}a/2)|<1$ for ZEG. It clearly shows the
well-known conclusion that a zero-energy edge state exists in the
region $k_{y}a \in ( 2\pi/3,4\pi/3)$ for ZEG. But the point of
$k_ya=\pm \pi$ must be excluded since
$t_{2}=2t\cos(k_{y}a/2)=2t\cos(\pi/2)=0$ that corresponds to an
internal mode within the isolated chains. From the conclusion of 1D
model in Sec.\ref{1Dsec}, we also can conclude that the zero-energy
edge state is stable.

\begin{figure}
\includegraphics[scale = 0.18]{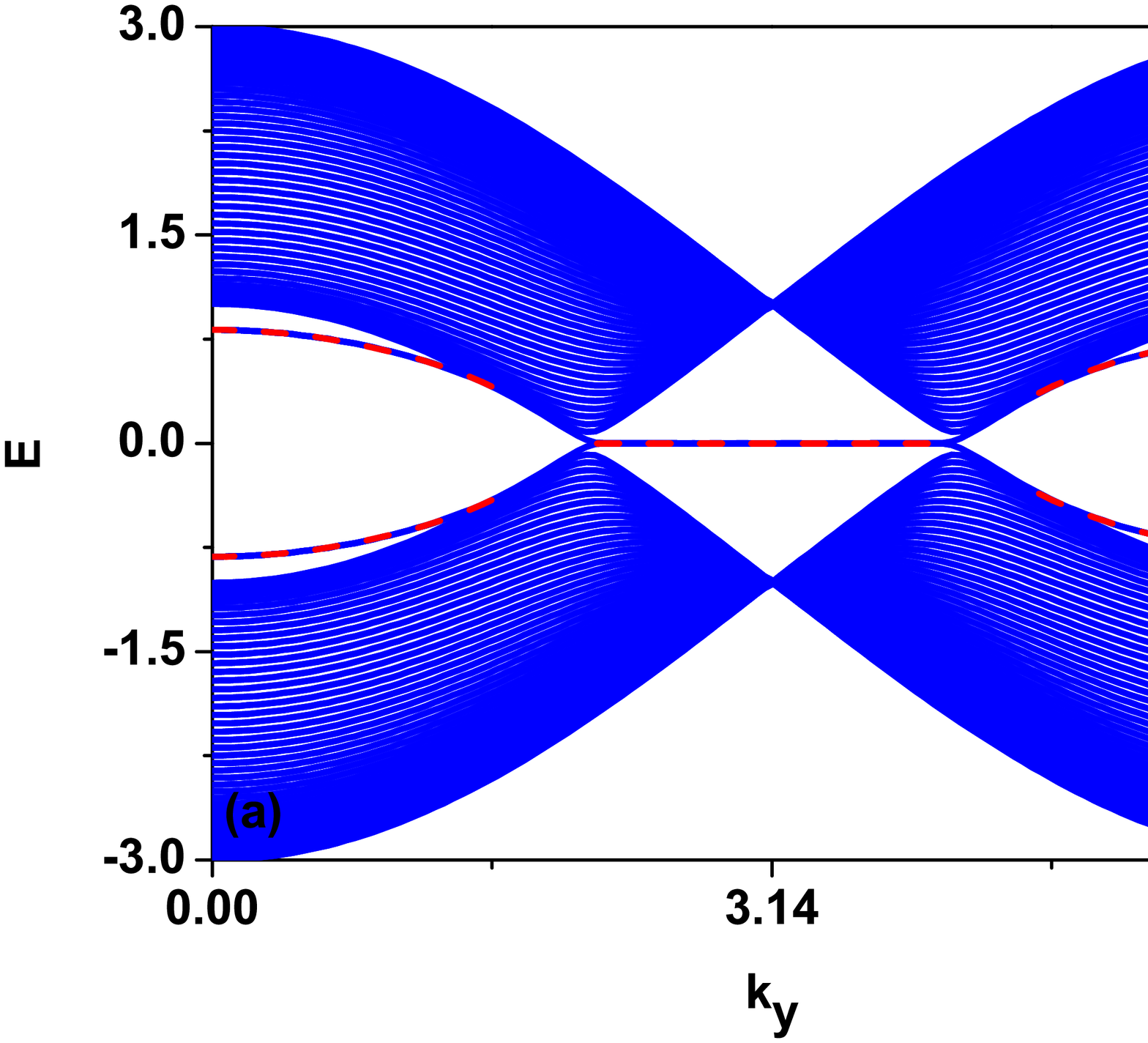} \qquad\quad
\includegraphics[scale = 0.18]{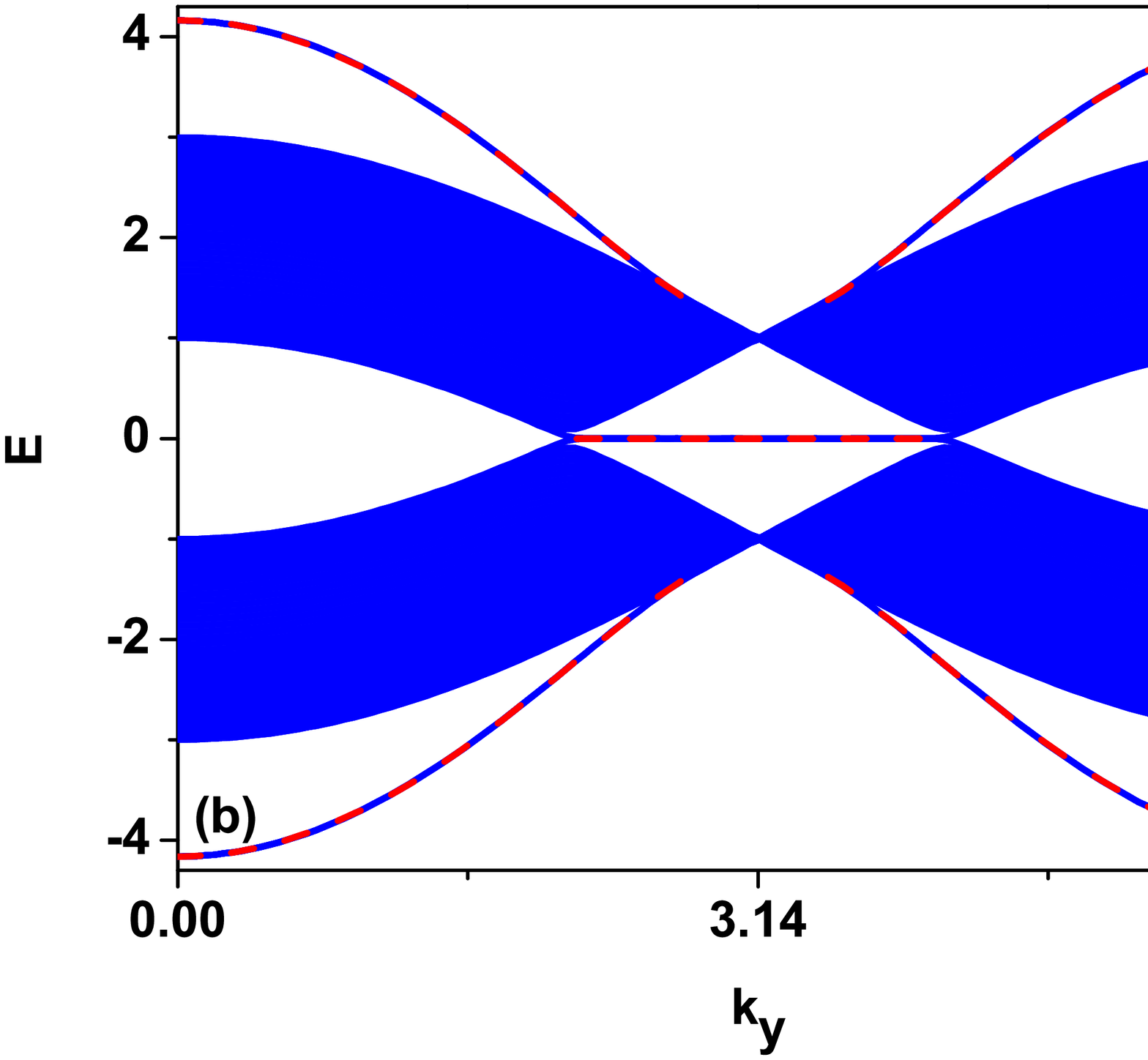}
\caption{(color online) By exact diagonalization method, the
dispersion relation of the bulk states and the edge state are
depicted (blue line), while the dispersion relation of the edge
state obtained through TMM is depicted in the dashed line (red
dash-line). Parameters: (a) $t'_{0}=0.5t$, (b) $t'_{0}=2.0t$.}
\label{ED-ana}
\end{figure}

Just as the discussion shown in the above section, the SNC for the
existence of the non-zero-energy edge states should be discussed
with $\alpha^{2}>4$. With similar discussions, we can get the
dispersion of energy spectrum of non-zero-energy edge states as
follows:
\begin{equation}\label{zigzagenergy}
E^{2} = 4
t'^{2}_{0}\cos^{2}{(k_{y}a/2)}+\frac{t'^{2}_{0}t^{2}}{t'^{2}_{0}-t^{2}}
\end{equation}
while the SNC for the existence of those states must satisfy:
\begin{equation}
\begin{cases}
t'_{0} < t, \; 0 < \frac{t^{2}}{2(t^{2}-t'^{2}_{0})}
                          < |\cos{(k_{y}a/2)}| \leqslant 1 \\
                \qquad\qquad \Longrightarrow t^{2} > 2 t'^{2}_{0};\\
\text{or}\\
t'_{0} > t, \; 0 < \frac{t^{2}}{2(t'^{2}_{0}-t^{2})}
                          < |\cos{(k_{y}a/2)}| \leqslant 1 \\
                \qquad\qquad \Longrightarrow t^{2} < 2 t'^{2}_{0}/3
\end{cases}
\end{equation}
As above shown, we can easily see that the existence of the
non-zero-energy edge states depends on the decoration on the edge,
the ratio between the edge hopping and the bulk hopping should be
within a certain threshold, so that we can detect the finite-energy
edge states. Comparing the results obtained above with those by the
means of the exact diagonal method on a finite-size zigzag
nanoribbon model, the results are accordance with each other, as
shown in Fig. \ref{ED-ana}.

\section{Conclusions}\label{conclusions}

We have studied the semi-infinite 1D lattice chain with Peierls
phase transition and ZEG model respectively, with the help of TMM.
We focus on the zero-energy edge states and non-zero-energy edge
states, under the condition of the edge decoration, which is the
electron hopping energy being different from that in bulk. In the 1D
model, we conclude that a zero-energy edge state will appear if
Peierls transition happens regardless edge decoration at all. A
non-zero-energy edge state can occur due to the edge decoration.
Then we map the 1D case into the ZEG. Suppose the electron hopping
energy is $t'_{0}$, different from that $t$ in bulk because of the
boundary effects. It is found that the existence of the zero-energy
edge states is independent of the magnitude of $t'_{0}$. In addition
to the zero-energy edge states, there can exist edge states of
finite energy if $t > \sqrt{2} t'_{0}$ or $t < \sqrt{2/3} t'_{0}$.
These edge states appear as pairs with energy $\pm E$ outside of the
continuum conduction and valence bands.

\acknowledgments We thank Dr. X. Z. Yan for helpful comments. This
work is supported by the National Natural Science Foundation of
China under grant No.10847001 and National Basic Research Program of
China (973 Program) under the grant (No.2009CB929204,
No.2011CB921803) project of China.

\end{document}